\documentstyle[sprocl,epsf]{article}

\def\be{\begin{equation}}
\def\ee{\end{equation}}
\def\bea{\begin{eqnarray}}
\def\eea{\end{eqnarray}}
\newcommand{\lt}{\left}
\newcommand{\rt}{\right}
\newcommand{\ov}{\overline}
\newcommand{\eq}[1]{(\ref{#1})}
\def\openone{\leavevmode\hbox{\small1\kern-3.8pt\normalsize1}}%

\newcommand{\nn}{\nonumber \\}
\newcommand{\no}{\nonumber }
\newcommand{\fig}[1]{Fig.~\ref{#1}}

\def\lsim{{~\raise.15em\hbox{$<$}\kern-.85em
          \lower.35em\hbox{$\sim$}~}}

\newcommand{\prd}{Phys.~Rev.~D}
\newcommand{\plb}{Phys.~Lett.~B}
\newcommand{\npb}{Nucl.~Phys.~B}

\newlength{\miniwidth}
\newlength{\miniwidthplot}
\newlength{\nseparation}
\newenvironment{nfigure}[1]
        {\begin{figure}[#1]\hrule\vspace{\nseparation}\par}
        {\vspace{\nseparation}\par \hrule \end{figure}}

\begin{document}

\title{Penguin contributions in the lifetime difference 
between $B_s$ and $B_d$, and a possible New Physics}

\author{ Yong-Yeon Keum \footnote{Talk given at
Fourth International Workshop on Particle Physics Phenomenology
at Kaohsiung, Taiwan, June 18-21, 1998},
\footnote{Monbushou Research Fellow} }
\address{Asia Pacific Center For Theoretical Physics \\
207-43 Cheongryangri Dong Dongdaemun-Gu, Seoul 130-012,
Korea}
\address{Theory Group, KEK, Tsukuba, Ibaraki 305-0801 Japan}


\maketitle\abstracts{
We consider penguin contributions to the lifetime splitting 
between the $B_s$ and the $B_d$ meson. In the Standard Model 
the penguin effects are found to be opposite in sign, but of 
similar magnitude as the contributions of the current-current 
operators, despite of the smallness of the penguin coefficients.  
We predict 
\begin{eqnarray}
\frac{\tau(B_s)}{\tau(B_d)} -1 &=& \lt( -1.2 \pm 10.0 \rt)
      \cdot 10^{-3} \cdot \lt(\frac{f_{B_s}}{190 \, \mathrm{MeV}} \rt)^2
, \no  
\end{eqnarray}
where the error stems from hadronic uncertainties.  Since penguin
coefficients are sensitive to new physics and poorly tested
experimentally, we analyze the possibility to extract them from a
future precision measurement of $\tau \lt(B_s \rt)/\tau\lt(B_d\rt)$.
Anticipating progress in the determination of the hadronic parameters
$\varepsilon_1$, $\varepsilon_2$ and $f_{B_s}/f_{B_d}$ we find that
the coefficient $C_4$ can be extracted with an uncertainty of order
$|\Delta C_4|\simeq 0.1$ from the double ratio
$[\tau(B_s)-\tau(B_d)]/[\tau(B^+)-\tau(B_d)]$, if
$|\varepsilon_1-\varepsilon_2|$ is not too small. }

\section{Introduction}
The theoretical achievement of the Heavy Quark Expansion (HQE)
\cite{hqe1} has helped a lot to understand the inclusive properties of
B-mesons. The comparision of the theoretical predictions with 
the experimental measurements for the heavy hadron lifetimes and 
their ratios is an important test of the theory of inclusive decays, 
 and the HQE at the order $(\Lambda_{QCD}/m_b)^3$ which is closely
related to the local quark-hadron
dualitiy, which is a priori assumption in inclusive non-leptonic decay.

When we neglect higher-order corrections in $1/m_b$, the semileptonic and
non-leptonic widths only depend on the CKM-factors and quarks masses.
This implies that all the heavy hadrons have the same lifetime and 
semileptonic width. Lifetime differences aries either from corrections
of ${\cal O}(1/m_b^2)$ originated by heavy-hadron ``wave-function'' effects,
 or from  ${\cal O}(1/m_b^3)$ corrections induced 
by non-spectator corrections; i.e., Pauli-interference and W-exchange
diagrams.
   
In the former case, the corrections to lifetime-universality are due
to the heavy-quark kninetic energy $\lambda_1$ and
the chromomagnetic term $\lambda_2$ which differ for different hadrons:
$\lambda_1(B_s) \neq \lambda_1(B)$ because of $SU(3)_f$ symmetry-breaking
effects, the mass of the strange qaurk is much larger than the mass of
 the u and d quarks; $\lambda_1(B) \neq \lambda_1(\Lambda_b)$
because mesons and baryons have different wave-functions; 
for the same reason the chromomagnetic term vanishes for the $\Lambda_b$ 
but not for the B meson.

The non-spectator contributions, although of ${\cal O}(1/m_b^3)$,
are enhanced by the factor $16 \pi^2$ due to the phase factor 
for $2 \rightarrow 2$ decay and for this reason may give sizeable effects.
Note that these corrections may have CKM factors different from those
of the leading terms.

By considering lifetime ratios, we can study most conveniently the lifetime
difference in B hadrons. In this way, one cancels the dependence on
quantities which are poorly known, such as the $b$-quark mass 
($\tau \propto m_b^5$),$|V_{cb}|$, and renormalons\cite{renormalon}.
  
\subsection{Experimental Measurement and Theoretical Prediction}
The average experimental results for lifetime ratios are \cite{f}:
\begin{equation}
{\tau(B_s) \over \tau(B^0)} = 0.98 \pm 0.07; \hspace{10mm}
{\tau(B^{\pm}) \over \tau(B^0)} = 1.07 \pm 0.04; \hspace{10mm}
{\tau(\Lambda_b) \over \tau(B^0)} = 0.78 \pm 0.04.
\label{Blifetime}
\end{equation}
The theoretical predictions is given as follows :
\begin{eqnarray}
&& \left|{\tau (B_s) \over \tau(B)} - 1 \right| < 1 \%   
\label{R1} \\
&& {\tau (B^{-}) \over \tau(B^0)} = 1 + 16 \pi^2 {f_B^2 M_B \over m_b^3}
\left[k_1 B_1 + k_2 B_2 + k_3 \epsilon_1 + k_4 \epsilon_2 \right]
\label{R2} \\
&& 
{\tau (B^{-}) \over \tau(B^0)} = 1 - 
{\lambda_1(\Lambda_b)-\lambda_1(B^0) \over 2 m_b^2} 
+ c_G {\lambda_2(\Lambda_b)-\lambda_2(B^0) \over  m_b^2} \nonumber \\
\cr
&& \hspace{10mm}+ 16 \pi^2 {f_B^2 M_B \over m_b^3}
\left[p_1 B_1 + p_2 B_2 + p_3 \epsilon_1 + p_4 \epsilon_2 
+ (p_5 + p_6 \tilde{B}) \right]
\label{R3} 
\end{eqnarray}
Using the experimental values of the hadron masses, hadronic parameters 
are given:
\begin{equation}
\lambda_1(B) - \lambda_1(\Lambda_b) = -(0.001 \pm 0.03) GeV^2, 
\label{R4}
\end{equation}
\begin{equation}
\lambda_2(B) \simeq 0.12 GeV^2, \hspace{10mm} 
\lambda_2(\Lambda_b) = 0.
\label{R5}
\end{equation}
\subsection{Hadronic paremeters}
When we define the Operator basis as follows
\begin{eqnarray}
{\cal O}_1 &=& \bar{b}\gamma_{\mu}(1 - \gamma_5) q \bar{q}\gamma^{\mu}
(1 - \gamma_5) b; \hspace{3mm}
T_1 = \bar{b}\gamma_{\mu}(1 - \gamma_5)T_a q \bar{q}\gamma^{\mu}
(1 - \gamma_5) T_a b; \\
\cr
{\cal O}_2 &=& \bar{b}(1 - \gamma_5) q \bar{q} (1 + \gamma_5) b; \hspace{3mm}
T_2 = \bar{b}(1 - \gamma_5)T_a q \bar{q} (1 + \gamma_5) T_a b; 
\end{eqnarray}
the matrix elements for B meson are defined in terms of their B-parameters
\cite{ns}: 
\begin{equation}
\langle B | {\cal O}_i | B \rangle  \equiv f_B^2 M_B^2 B_{i}(\mu); 
\hspace{10mm}
\langle B | {T_i} | B \rangle  \equiv f_B^2 M_B^2 \epsilon_{i}(\mu).
\end{equation} 
For B-baryon, the matrix elements are defined by \cite{ns,cheng}
\begin{equation}
\langle \Lambda_b | \bar{b}_{\alpha} \gamma^{\mu}(1 - \gamma_5)q_{\beta}
\bar{q}_{\beta} \gamma_{\mu}(1 - \gamma_5)b_{\alpha} \rangle
\equiv - \tilde{B} \langle \Lambda_b | {\cal O}_1 | \Lambda_b \rangle; 
\end{equation}
\begin{equation}
\langle \Lambda_b | {\cal O}_1 | \Lambda_b \rangle 
= - {f_{B_q}^2 m_{B_q}^2 \over 6} r ;
\end{equation}
We have to introduce 4 unknown parameters for B meson and 2 unknown parameters
for B baryon, which contain non-purtubative property.  
Now let us consider the penguin contributions to the lifetime
splitting between the $B_s$ and the $B_d$ meson.

\begin{nfigure}{tb}
\begin{minipage}[t]{0.48\textwidth}
\centerline{\epsfxsize=0.8\textwidth \epsffile{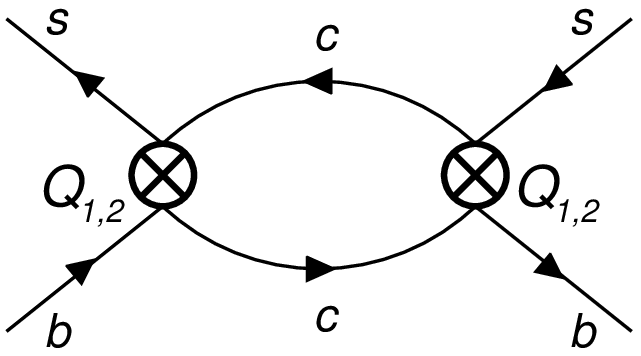}}
\caption{Non-spectator \emph{(weak annihilation)} contribution to the 
  $B_s$ decay rate involving two current-current operators. The
  corresponding diagram for the $B_d$ decay is obtained by replacing
  $s$ by $d$ and the upper $c$ by $u$.}\label{fig:cc}
\end{minipage}\hspace{2ex}
\begin{minipage}[t]{0.48\textwidth}
\centerline{\epsfxsize=0.8\textwidth \epsffile{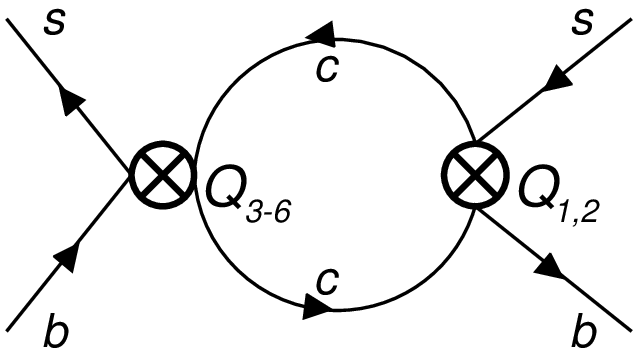}}
\caption{Weak annihilation diagram involving one penguin operator 
$Q_{3-6}$. Penguin contributions to the non-spectator rate of the 
$B_d$ meson are CKM suppressed and therefore negligible.  
}\label{fig:cp}
\end{minipage}
\end{nfigure}

\section{Penguin Contributions}
For the non-spectator contributions to the $B_s$ decay rate we need
the $|\Delta B|=|\Delta S|=1$-hamiltonian:
\begin{eqnarray}
  H &=& \frac{G_F}{\sqrt{2}} V_{cb} V_{cs}^* 
        \left[ \sum_{j=1}^6 C_j Q_j + C_8 Q_8 \right] \; 
\label{hd}
\end{eqnarray}
with 
\begin{eqnarray}
Q_1 \; = \; (\bar{s}^{\alpha}c^{\beta})_{V-A}  
\cdot (\bar{c^{\beta}}b^{\alpha})_{V-A},
&& \qquad
Q_2 \; = \; (\bar{s}c)_{V-A}  \cdot (\bar{c}b)_{V-A} \no \\[2mm]
Q_3 \; = \; \!\!\!\!\!\!\!
\sum_{q=u,d,s,c,b} (\bar{s}b)_{V-A}  \cdot (\bar{q}q)_{V-A}   ,
 && \qquad Q_4 \; = \; \!\!\!\!\!\!\! 
         \sum_{q=u,d,s,c,b} (\bar{s}^{\alpha}b^{\beta})_{V-A}  
                       \cdot (\bar{q}^{\beta}q^{\alpha})_{V-A} 
\no \\[2mm] 
Q_5 \; = \; \!\!\!\!\!\!\!
 \sum_{q=u,d,s,c,b} (\bar{s}b)_{V-A}  \cdot (\bar{q}q)_{V+A},
 && \qquad Q_6 \; = \; \!\!\!\!\!\!\! \sum_{q=u,d,s,c,b} 
(\bar{s}^{\alpha}b^{\beta})_{V-A}  \cdot 
(\bar{q}^{\beta}q^{\alpha})_{V+A}
  \no \\[2mm]
 Q_8 \; =\;  - \frac{g}{8 \pi^2} \, m_b \,\bar{s} \sigma^{\mu \nu} 
                 \lt( 1+\gamma_5 \rt)  T^a b \cdot G^a_{\mu \nu} 
 \; . \hspace{-6em}
\label{basis}
\end{eqnarray} 
In \eq{hd} we have set $V_{ub} V_{us}^* =
O(\lambda^4)$ to zero.  The diagram of \fig{fig:cc} has been
calculated in \cite{ns,bbd} and yields contributions to the
non-spectator part $\Gamma^{\mathrm{non-spec}}$ of the $B_s$ decay
rate proportional to $C_2^2, C_1 \cdot C_2$ and $C_1^2$. 
$\tau(B_s)/\tau(B_d)-1$ is proportional to $\Gamma^{\mathrm{non-spec}}
(B_d)-\Gamma^{\mathrm{non-spec}} (B_s)$.  The main differences between
the result of \fig{fig:cc} for these two rates are due to the
different mass of $u$ and $c$ and the difference between $f_{B_d}$ and
$f_{B_s}$. Hence the current-current parts of $\tau(B_s)/\tau(B_d)-1$
proportional to $C_2^2, C_1 \cdot C_2$ or $C_1^2$ are suppressed by a
factor of $z$ or $\Delta$ with
\begin{eqnarray}
z \; = \; \frac{m_c^2}{m_b^2} \; = \; 0.085 \pm 0.023 ,
\qquad 
\Delta \; = \; 1- \frac{ f_{B_d}^2 M_{B_d} }{ f_{B_s}^2 M_{B_s} } 
\; = \; 0.23 \pm 0.11 . \label{latt}
\end{eqnarray}
The result for $\Delta$ in \eq{latt} is the present world average of
lattice calculations \cite{l}. There are also $SU(3)_F$ violations in
the B-factors, but they are expected to be small from the experience
with those appearing in $B^0-\ov{B}{}^0$-mixing. 
In this analysis we use the same $B_1$, $B_2$, $\epsilon_1$, and
$\epsilon_2$ in $\tau(B_s)$ and $\tau(B_d)$.

\begin{nfigure}{tb}
\begin{minipage}[t]{0.48\textwidth}
\centerline{\epsfxsize=0.9 \textwidth \epsffile{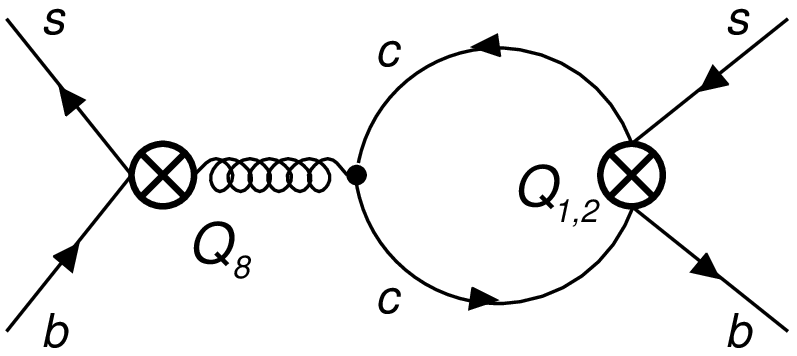}}
\caption{Contribution of $Q_8$ to $\Gamma^{\mathrm{non-spec}}(B_s)$. 
In the Standard Model the diagram is of the same order of magnitude 
as radiative corrections to \fig{fig:cp} and therefore negligible. 
Yet in models in which quark helicity flips occur in  flavour-changing 
vertices $|C_8|$ can easily be ten times larger than in the Standard 
Model. The contribution of $Q_1$ vanishes.}\label{fig:8}
\end{minipage} \hspace{2ex}
\begin{minipage}[t]{0.48\textwidth}
\centerline{\epsfxsize=0.9\textwidth \epsffile{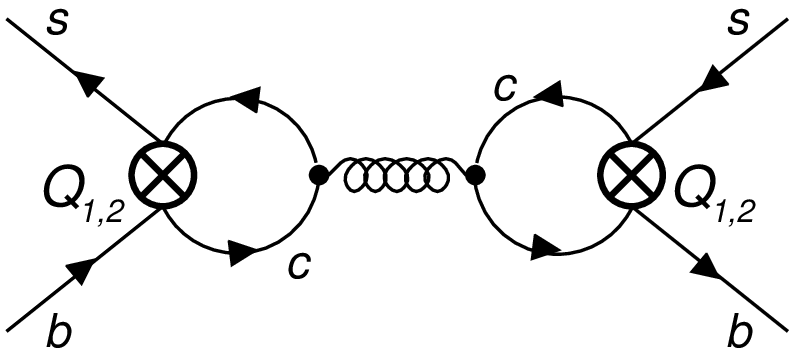}}
\caption{Penguin diagram contribution to
  $\Gamma^{\mathrm{non-spec}}(B_s)$. The final state corresponds
  to a cut through either of the $(\ov{c},c)$-loops. The contributions
  of $Q_1$ vanish by colour. This is the only NLO contribution to 
  $\tau(B_s)/\tau(B_d)-1$ involving $Q_{1,2}$ without suppression 
  factors of $\Delta$ or $z$. } \label{fig:nlo}
\end{minipage}
\end{nfigure}

Our result for the non-spectator part of the $B_c$ decay rates  
reads:
\begin{eqnarray}
\Gamma^{\mathrm{non-spec}} \left( B_s\right)
     &=& - \frac{G_F^2 m_b^2}{12 \pi} 
        \left| V_{cb} V_{cs} \right|^2 \sqrt{1- 4 z} f_{B_s}^2 M_{B_s}
 \nonumber \\ 
     & & \left[ a_1 \, \varepsilon_1 + a_2 \, \varepsilon_2 + 
             b_1 B_1 + b_2 B_2 \right] \label{dgbs}
\end{eqnarray}
with 
\begin{eqnarray}
a_1 &=& \left[ 2 C_2^2 + 4 C_2 C_4^\prime  \right] 
        \left[1 - z \right] + 
        12 z C_2 C_6^\prime + \left[ 1 + 2 z  \right] \frac{\alpha_s}{\pi} 
        C_2 C_8 \nonumber \\ 
a_2 &=& - \left[ 1 + 2 z \right] \left[ 2 C_2^2 + 4 C_2 C_4^\prime  +
               \frac{\alpha_s}{\pi} C_2 C_8 \right] \nonumber \\ 
b_1 &=&  \left[ C_2 + N_c C_1  \right]
        \left\{ \left( 1 - z  \right) 
     \left[  \frac{C_2}{N_c}  +  C_1   + 
              2 C_3^\prime + 2 \frac{C_4^\prime}{N_c} \right]
                    + 6 z \left[ C_5^\prime + \frac{C_6^\prime}{N_c} \right] 
        \right\} \nonumber \\
b_2 &=& - \left[ 1 + 2 z  \right] \left[ C_2 + N_c C_1 \right] \left\{ 
        \frac{1}{N_c} \left[ C_2 + N_c C_1 \right] + 
      2 \left[ C_3^\prime + \frac{C_4^\prime}{N_c} \right]
        \right\} \label{spc}
\end{eqnarray}
Here $N_c=3$ is the number of colours. By setting $C_j^\prime$,
$j=3,\dots,6$, and $C_8$ in \eq{spc} to zero one recovers the result
of Ref.[4].\footnote{Notice that our notation of $C_1$ and $C_2$ is
  opposite to the one in Ref.[4].}
The result for the non-spectator contributions to the $B_d$ decay
rate reads \cite{ns}:
\begin{eqnarray}
\!\!\!\!\!
\Gamma^{\mathrm{non-spec}} \lt( B_d\rt) \!\!\!
&=& \!\!\! \frac{G_F^2 m_b^2}{12 \pi} 
        \lt| V_{cb} V_{ud} \rt|^2 \lt(1- z\rt)^2 f_{B_s}^2 M_{B_s} 
        \lt( \Delta -1 \rt) \nn 
&&        \lt[ a_1^d \, \varepsilon_1 + a_2^d \, \varepsilon_2 + 
             b_1^d B_1 + b_2^d B_2 \rt] \label{nsd}
\end{eqnarray}
with\footnote{In the large $N_c$ limit one finds
  $\Gamma^{\mathrm{non-spec}}$ helicity suppressed in analogy to the
  leptonic decay rate. This shows that one cannot neglect the
  $O(1/N_c)$ terms.}
\begin{eqnarray}
&&
\begin{array}[b]{rclrcl}
\displaystyle
a_1^d &=& \displaystyle 2 C_2^2 \lt( 1+ \frac{z}{2} \rt), \qquad &
\displaystyle
a_2^d &=& \displaystyle - 2 C_2^2 \lt( 1 + 2 z  \rt), \\
\displaystyle
b_1^d &=& \displaystyle \frac{1}{N_c} \lt( C_2 + N_c C_1 \rt)^2  
            \lt( 1 + \frac{z}{2} \rt), \qquad &
\displaystyle
b_2^d &=& \displaystyle - \frac{1}{N_c}
          \lt( C_2 + N_c C_1 \rt)^2  \lt( 1 + 2 z \rt) . 
\end{array} 
\label{specd}
\end{eqnarray}
When we combine (\ref{dgbs}-\ref{specd}) in order to predict\cite{uli-keum} 
$\tau (B_s)/\tau (B_d) -1$ : 
\begin{eqnarray}
\frac{\tau (B_s)}{\tau (B_d)} -1 &=& 
    \frac{\Gamma^{\mathrm{non-spec}}(B_d) - 
          \Gamma^{\mathrm{non-spec}}(B_s) }{ \Gamma^{\mathrm{total}} } 
+O (10^{-3})
\cr
&=& 9.0 \cdot 10^{-4} \epsilon_1 - 1.63 \cdot 10^{-2} \epsilon_2
+ 2.0 \cdot 10^{-4} B_1 - 5.0 \cdot 10^{-4} B_2
\end{eqnarray}

The experimental world average \cite{f}
\begin{eqnarray}
\frac{\tau (B^+)}{\tau (B_d^0)} &=& 1.07 \pm 0.04 \label{expd}
\end{eqnarray}
leads to the following constraint:
\begin{eqnarray}
  \varepsilon_1 &\simeq & \lt( -0.2 \pm 0.1 \rt) \lt( \frac{0.17
  \,\mathrm{GeV}}{f_B} \rt)^2 \lt( \frac{m_b }
{4.8 \,\mathrm{GeV}} \rt)^3 + 0.3
  \varepsilon_2 + 0.05 . \label{con}
\end{eqnarray} 
 Here we consider the range
$|\varepsilon_1|$, $|\varepsilon_2|\leq 0.3$, and further obey \eq{expd}.

From Tab.2 of ref [8] we realize that the penguin contributions
 are comparable in size, but opposite in sign 
to the current-current part obtained in Ref.[4]. 
This makes the experimental detection of any deviation of 
$\tau(B_s)/\tau(B_d)$ from 1 even more difficult, 
if the penguin coefficients are really
dominated by Standard Model physics. 
From the results of Tab.2 in Ref.[8], we obtain 
\begin{eqnarray}
\frac{\tau(B_s)}{\tau(B_d)} -1 &=& \lt( -1.2 \pm 8.0 \pm 2.0 \rt)
\cdot 10^{-3} \cdot \lt(\frac{f_{B_s}}{190 \, \mathrm{MeV}} \rt)^2
\lt( \frac{4.8 \,\mathrm{GeV}}{m_b } \rt)^3 . \label{nums}
\end{eqnarray}
Here the first error stems from the uncertainty in $\varepsilon_1$ and
$\varepsilon_2$ and will be reduced once lattice results for the
hadronic parameters are available. The second error summarizes the
remaining uncertainties.

\section{New Physics Effects}
Today we have little experimental information on the sizes of the
penguin coefficients. Their smallness in the Standard Model allows for
the possibility that they are dominated by new physics.  The total
charmless inclusive branching fraction $Br(B \rightarrow \textit{no
  charm})$ is a candidate to detect new physics contributions to $C_8$
\cite{k}, but it is much less sensitive to $C_{3-6}$ \cite{lno}. 
Eq.(17) of Ref.[8]  reveals that
$\tau(B_s)/\tau(B_d)$ is a complementary observable mainly sensitive
to $C_4 $, while $C_8$ is of minor importance. 

 New physics contributions $\Delta
C_{3-6} ( \mu = 200 GeV) $ affect $C_4 ( \mu = 4.8\, \mathrm{GeV})$ by
\begin{eqnarray}
\Delta C_4 ( \mu = 4.8\, \mathrm{GeV}) &=& 
-0.35 \, \Delta C_3 ( 200 \, \mathrm{GeV}) + 
 0.99 \, \Delta C_4 ( 200  \, \mathrm{GeV})  \nn
&& -0.03 \, \Delta C_5 ( 200 \, \mathrm{GeV})  
 -0.22 \, \Delta C_6 ( 200 \,  \mathrm{GeV}) .
\end{eqnarray}   
Observe that $\Delta C_4 (200 \, \mathrm{GeV})=-0.05$ already increases
$C_4^\prime (m_b)$ by more than a factor of two. 

Clearly the usefulness of $\tau(B_s)/\tau(B_d)$ to probe $C_{3-6}$
crucially depends on the size of $|\varepsilon_1-\varepsilon_2|$ and
$f_{B_s}$. We now investigate the sensitivity of $\tau(B_s)/\tau(B_d)$
to $\Delta C_4 (\mu = m_b)$ in a possible future scenario for 
the hadronic parameters. We assume
\begin{eqnarray}
\varepsilon_1 \; = \; -0.10 \pm 0.05, \quad 
\varepsilon_2 \; = \; 0.20 \pm 0.05,  \quad
B_1,B_2 \; = \; 1.0 \pm 0.1, && \nn
f_{B_s} \; = \; \lt(190 \pm 15 \rt) \, \mathrm{MeV}, \quad 
\Delta \;  = \; 0.23 \pm 0.05, \quad
m_b \; = \; \lt( 4.8 \pm 0.1 \rt) \, \mathrm{GeV}.&& 
\label{sc}
\end{eqnarray}

A cleaner observable which can see a new physics effects is
the double ratio:
\begin{eqnarray}
\frac{\tau(B_s)-\tau(B_d)}{\tau(B^+)-\tau(B_d)} &=& 
\frac{B_{SL}(B_s)-B_{SL}(B_d)}{B_{SL}(B^+)-B_{SL}(B_d)} ,
 \label{dr}
\end{eqnarray} 
which depends on $\varepsilon_1, \varepsilon_2$  and $\Delta$, while the 
dependence on $f_{B}$ and $m_b$ cancels. The corresponding plot for 
the parameter set of \eq{sc} can be found in \fig{fig:plot2}
 If $\Delta C_4 < -0.075$ or $\Delta C_4 >
0.140$, we find the allowed range for
$[\tau(B_s)-\tau(B_d)]/[\tau(B^+)-\tau(B_d)]$ incompatible with the
Standard Model.
 The  experimental detection of a sizeable negative lifetime difference
$\tau(B_s)-\tau(B_d)$ may reveal non-standard contributions to
$C_4^\prime$ of similar size as its Standard Model value.
\fig{fig:plot2} shows that e.g.\ the bound $\tau(B_s)-\tau(B_d)< -0.20
[\tau(B^+)-\tau(B_d)]$ would indicate $\Delta C_4 > 0.051$.  
We conclude that the detection of new physics contributions to $C_4$ of
order $0.1$ is possible with precision measurements of
$\tau(B_s)/\tau(B_d)$. 
 
\section{Conclusions}
We have calculated the contributions of the penguin operators
$Q_{3-6}$, of the chromomagnetic operator $Q_8$ and of penguin
diagrams with insertions of $Q_2$ to the lifetime splitting between
the $B_s$ and $B_d$ meson. In the Standard Model the penguin effects
are found to be roughly half as big as the contributions from the
current-current operators $Q_1$ and $Q_2$, despite of the smallness of
the penguin coefficients. Yet they are opposite in sign, so that any
deviation of $\tau(B_s)-\tau(B_d)$ from zero is even harder to detect
experimentally.  Assuming a reasonable progress in the determination
of the hadronic parameters a precision measurement of
$\tau(B_s)/\tau(B_d)$ can be used to probe the coefficient $C_4$ with
an accuracy of $|\Delta C_4|=0.1$. Hence new physics can only be
detected, if $C_4$ is dominated by non-standard contributions.  The
sensitivity to $C_4$ depends crucially on the difference of the
hadronic parameters $\varepsilon_1$ and $\varepsilon_2$.  For the
extraction of $C_4$ the double ratio
$[\tau(B_s)-\tau(B_d)]/[\tau(B^+)-\tau(B_d)]$ turns out to be very
useful to see a new physics effects. 
\begin{nfigure}{tb}
\centerline{\epsfxsize=0.9\textwidth \epsffile{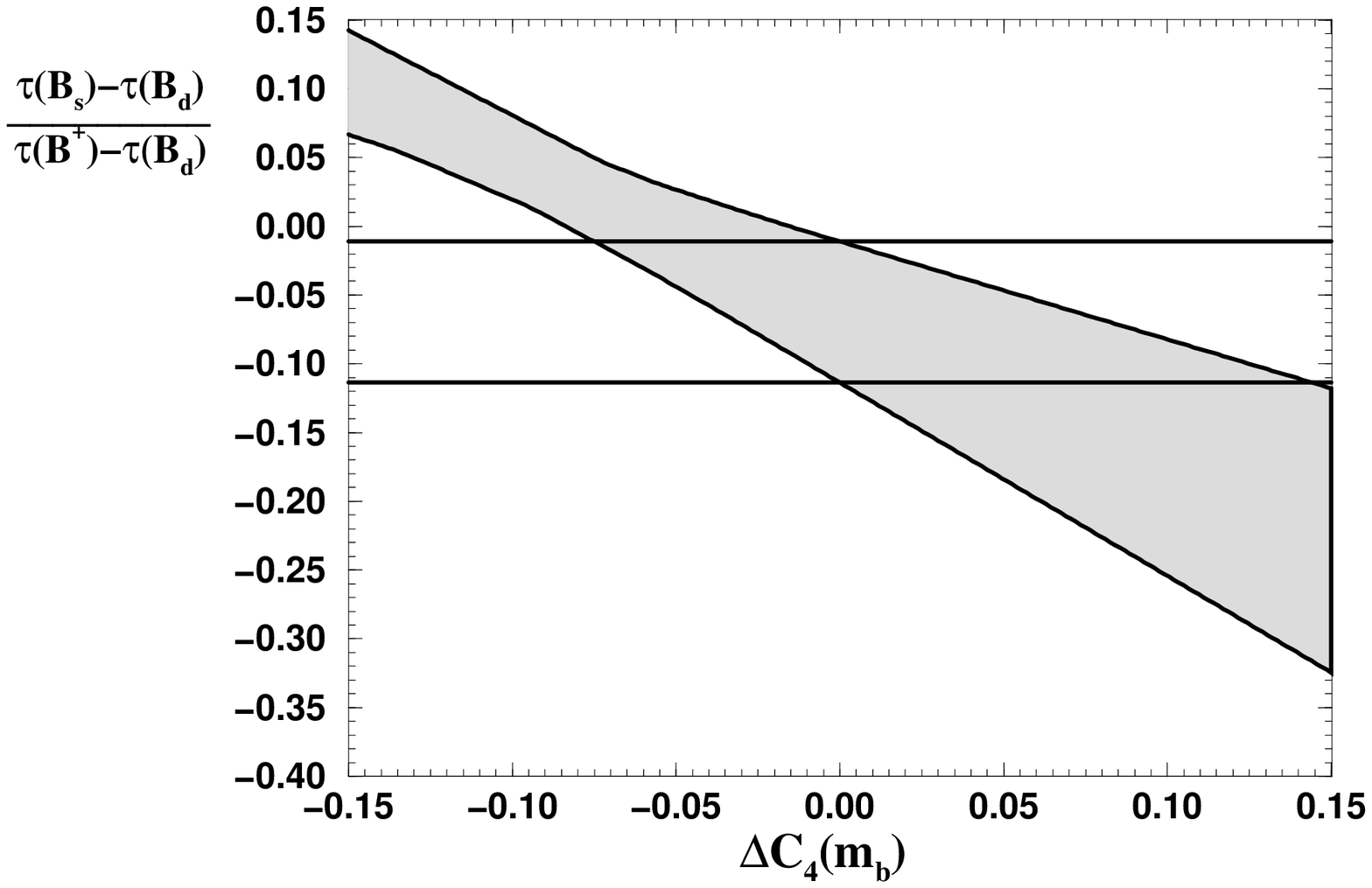}}
\caption{Dependence of $(\tau(B_s)-\tau(B_d))/(\tau(B^+)-\tau(B_d))$ 
  on $\Delta C_4$ for the parameter set in \eq{sc}. This double ratio
  depends on $f_{B_s}$ and $f_{B_d}$ only through $\Delta$, and the 
  factor of $m_b^{-3}$ cancels. 
}\label{fig:plot2}
\end{nfigure}
\section*{Acknowledgements}
The author would like to thank the organizers of the fourth international
 workshop on particle physics phenomenology at Kaoshiung.
Y.-Y.Keum thanks Chris Sachrajda and Hai-Yang Cheng
 for helpful discussions and U. Nierste for enjoyable collaboration
on this subject.  He is grateful to Prof. M. Kobayashi and
Prof.\ W.~Buchm\"uller for their hospitality and encouragement.
 This work is supported
in part by the Basic Science Research Institute Program,
Ministry of Education, Project No. BSRI-97-2414,
and in part by the Grant-in Aid for Scientific
from the Ministry of Education, Science and Culture, Japan.

\end{document}